\documentclass[aps,reprint,prl,groupedaddress]{revtex4-1}
\usepackage{amsmath,bm}
\usepackage{epsfig}
\usepackage{hyperref}
\usepackage{setspace}
\usepackage{color}

\def\uu{{\bm u}}
\def\xx{{\bm x}}
\begin{document}
\sloppy

\title{How vortices and shocks provide for a flux loop in two-dimensional \\compressible turbulence}

\author{Gregory Falkovich$^{1,2}$} \email{gregory.falkovich@weizmann.ac.il}
\author{Alexei G. Kritsuk$^3$}\email{akritsuk@ucsd.edu}

\affiliation{$^1$Weizmann Institute of Science, Rehovot 76100, Israel\\
$^2$Institute for Information Transmission Problems, Moscow 127994, Russia\\
$^3$University of California, San Diego, 9500 Gilman Drive, La Jolla, California 92093-0424, USA }

\begin{abstract}
Large-scale turbulence in fluid layers and other quasi-two-dimensional  compressible systems consists of planar vortices and waves. Separately, wave turbulence usually produces a direct energy cascade, while solenoidal planar turbulence transports energy to large scales by an inverse cascade. Here, we consider turbulence at finite Mach numbers when the interaction between acoustic waves and vortices is substantial. We employ solenoidal pumping at intermediate scales and show how both direct and inverse energy cascades are formed starting from the pumping scale. 
We show that there is an inverse cascade of kinetic energy up to a scale $\ell$, where a typical velocity reaches the speed of sound; this creates shock waves, which provide for a compensating direct cascade. When the system size is less than $\ell$, the steady state contains a system-size pair of long-living condensate vortices connected by a system of shocks. Thus turbulence in fluid layers processes energy via a loop: Most energy first goes to large scales via vortices and is then transported by waves to small-scale dissipation.
\end{abstract}


\maketitle

Inverse  cascade is  a counterintuitive process of self-organization of turbulence.
Predicted almost simultaneously for two-dimensional (2D) incompressible flows \cite{Kra} and  sea wave turbulence \cite{Zakharov} and established in many cases of turbulence in plasma, optics, etc. \cite{Tab,Shats,TFMK,IGW,Opt,Cond}, it is predicated on the existence of two quadratic conserved quantities having different wave-number dependencies. Excitation at some intermediate wave number then leads to two cascades: a direct one to small scales and an inverse one to large scales.
There is always a strong dissipation at small scales which acts as a sink for the direct cascade. On the contrary, large-scale motions are usually less dissipative, so that an inverse cascade can proceed unimpeded, either producing larger and larger scales or reaching the box size and creating a coherent mode of growing  amplitude. That process is now actively studied in 2D incompressible turbulence \cite{Tab,Shats,Axel,laurie14,F16,KL16,FFL17}, including in a curved space, where vortex rings rather than vortices are created \cite{FG}.
The energy of an incompressible flow in an unbounded domain grows unlimited when the friction factors go to zero at a finite energy input rate. The same happens to the action of wave turbulence \cite{ZLF}, if long waves of large amplitude are stable.
For example, optical turbulence in media with defocusing nonlinearity produces a growing condensate \cite{Opt,Cond,VDF}. On the contrary,  in the focusing case,  condensate instability results in wave collapses which provide for a loop of inverse cascade to the small-scale dissipation so that there is a steady state with only small-scale dissipation \cite{Opt}.

Here, we consider compressible two-dimensional turbulence 
which is of significant importance for numerous geophysical, astrophysical and industrial applications. We show that it realizes a third possibility of a steady state with only small-scale dissipation: On the one hand, an inverse cascade is able to produce long-living stable vortices, and on the other hand, the system reaches a steady state as the vortices produce waves that break and dissipate the energy. Two-dimensional compressible hydrodynamics describes motions in fluid layers on scales exceeding the fluid depth when vortices are planar while waves are acoustic,  the thickness playing the role of density. We consider an ideal-gas equation of state with the ratio of specific heats $\gamma=c_p/c_v\to1$ (that is 
 near isothermal) which is relevant to astrophysical systems (where radiation provides for temperature equilibration \cite{suppl,wolfire...03,field65,kritsuk86,krasnobaev.17,kritsuk..17,mckee.07,martos.98}) and for soap films flows with a large Reynolds number, a nonvanishing Mach number, and negligible solubility \cite{Chomaz01,fast05}. 

 Our results may also relate to the shallow water model, which is basically described by the same set of equations, but with $\gamma=2$. In this context, some of the potential applications include dissipation in geostrophic turbulence and its impact on the stability of mesoscale oceanic eddies \cite{mcwilliams16}. A weak direct energy cascade was predicted in Ref.~\cite{warn86}, using statistical mechanical arguments, which seemed in contradiction to the numerical results of Ref.~\cite{farge.88}. A flux loop of a similar nature to the one discussed here perhaps might solve that paradox.

\begin{figure*}
	\centering
	\includegraphics[scale=.6]{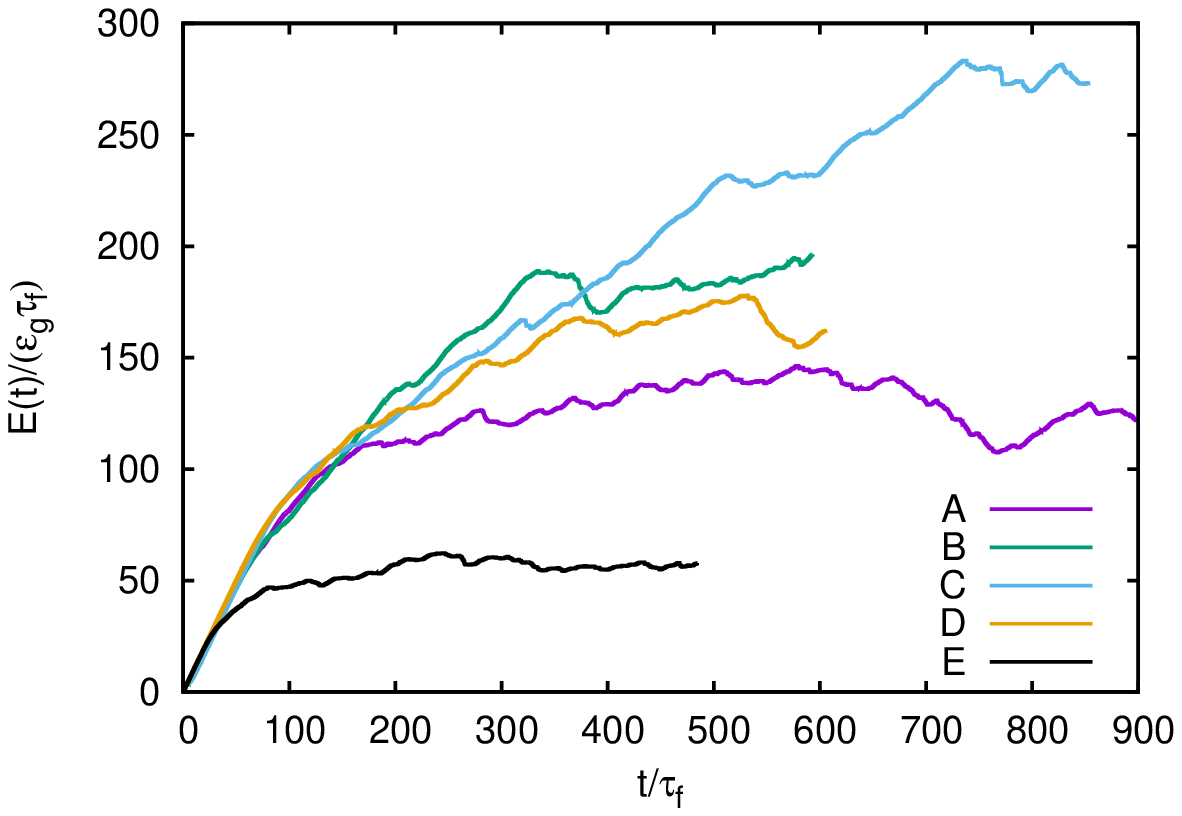}
	\includegraphics[scale=.6]{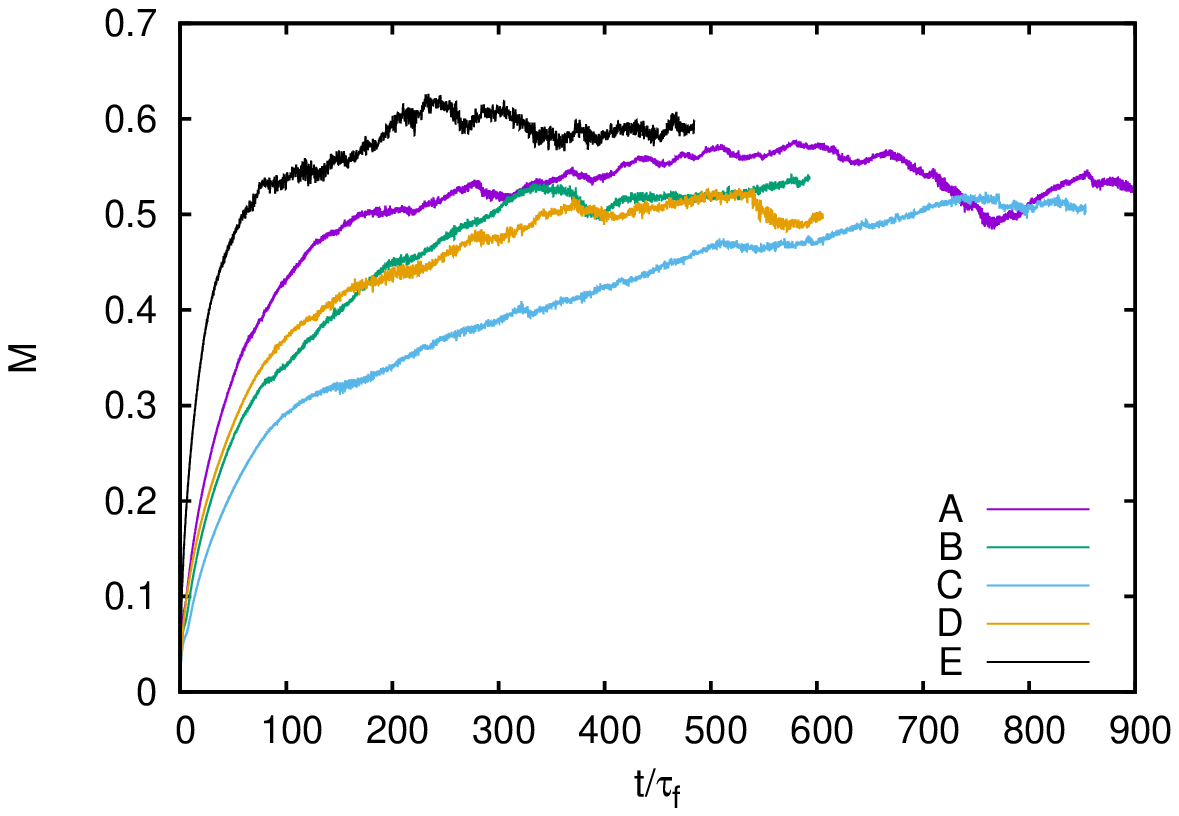}
	\caption{Time evolution of the total energy (left) and the rms Mach number (right) for cases A---E. The time is normalized with the forcing time $\tau_f\equiv\rho_0^{1/3}\lambda_f^{2/3}\varepsilon_f^{-1/3}$. Cases A---C form a sequence with increasing grid resolution, where each step adds a factor of 2 to the extent of both cascades in this dual-cascade setup.}
	\label{e-of-t}
\end{figure*}

A flux loop was also observed in a relatively simpler case of weakly stratified 2D turbulence \cite{Strat}, where there is only one conserved quantity, so that an inverse cascade can exist only in a restricted interval of scales until the kinetic energy  converts into the potential energy which cascades to small scales. Our case is more complicated and rich: First, because the energy, 
$E=K+W=\int [\rho u^2/2+c^2\rho\ln(\rho/\rho_0)]d\xx$, written here for an isothermal case ($\rho$ is the density, $\rho_0$ the mean density,  $\bm u$ the velocity, $c$ the sound speed), 
is both potential and kinetic, and also because the kinetic energy has two components, solenoidal and potential (dilatational). 
Second,  smooth flows in ideal 2D compressible hydrodynamics conserve not only the energy integral 
but also the potential vorticity $\bm\omega/\rho$ of any streamline, where $\bm\omega=\bm\nabla\times{\bm u}$.  We characterize compressibility by the rms Mach number $M=\sqrt{\langle u^2\rangle}/c$.
When compressibility is small, two cascades exist, much as in the incompressible case \cite{D}. Indeed, the two relevant conserved quantities are close to quadratic: (i) The density times the squared potential vorticity,  $H=\int  \omega^2/\rho\, d\xx$, goes into the direct cascade, and (ii) the (mostly kinetic) energy
goes into the inverse cascade. As the vortices get larger and faster in the inverse cascade, they start to create density perturbations thus increasing the potential energy along with the kinetic energy. Even when external pumping is weak, as the inverse cascade proceeds to larger scales, typical velocities increase and eventually become comparable to the speed of sound, while density perturbations become substantial. That allows for an effective interaction of vortices and waves and energy transfer from the former to the latter. Waves can then transfer energy back from large to small scales due to wave breaking and shock creation. We show that  kinetic energy has an upscale flux above the force scale $\lambda_f$. 
Since we observe a steady state, then the return downscale flux must be of potential energy. What is remarkable is that the fluxes are independent of wave number $k$ at $k<k_f\equiv2\pi/\lambda_f$, thus representing cascades. 

The description of numerical simulations  (implicit LES \cite{sytine....00,grinstein..07,ppm,enzo}) can be found in Ref.~\cite{suppl}.
Before analyzing the steady state, let us describe the energy growth, saturation, and fluctuations.
At small $t$, while $M<0.2$, the kinetic part strongly dominates the energy balance. On average, the contribution of potential  energy reaches $W\approx0.1E$ at $M=0.5$--$0.7$, as the total energy growth saturates. Remarkably, this $10$\% saturation level does not depend on the pumping rate $\varepsilon_f$. Simulations show that  $K(t)$ and $W(t)$ are strongly coupled and oscillate with opposite phases. The oscillation amplitude grows with $M$ and eventually saturates, reaching $W\approx0.15E$ during dissipation bursts and decreasing to $0.07E$ during periods of more quiet evolution. The main characteristic frequency of such oscillations is determined by the sound speed $c$, rotation velocity profile of large-scale vortices $U(r)$, and mean intervortical separation $L/\sqrt{2}$ ($L$ is the domain size). These large-scale acoustic oscillations represent a compressible component of the total condensate energy.

Figure~\ref{e-of-t} presents the evolution of energy and Mach number for the cases of weak (A, B, C), intermediate (D), and strong (E) pumping $\varepsilon_f$. In all cases, the energy evolution starts with a linear growth $E(t)=\varepsilon_ft$, which soon saturates due to a strong peak of small-scale compressible dissipation. After that, the inverse cascade is launched and linear growth resumes with a lower rate $\dot{E}=\varepsilon_g\sim0.9\varepsilon_f$. As soon as the rms Mach number exceeds 0.3, shock dissipation starts to play a role and further slows down the energy growth to $\sim0.4\varepsilon_f$. Somewhat later, a domain-size vortex dipole appears and then grows more coherent, as more energy is pumped in. The periods of slow growth are interrupted by episodic bursts of shock dissipation. For instance, the emergence of a single strong shock (pressure jump $>2$), connecting the centers of large-scale vortices can cause a sharp drop in $E(t)$ by several percent.  As the bursts become more frequent at higher energy levels ($M>0.5$), the overall growth slows down to below 10\% and eventually saturates, but one can still see short patches of uninterrupted growth with the same characteristic rate of $0.4\varepsilon_f$ between the bursts. Bursts of dissipation are observed to be of a different nature. Some are correlated with the intermittent appearance of shocks across the large-scale vortex dipole that become particularly dramatic when the two vortices approach closely. Some other cases correspond to vortices being destroyed completely and then reappearing after a while.
Deep energy minima  are accompanied by intense oscillations of relative strength of the vortices in the pair as best seen in the density movie \cite{suppl}; apparently, the large-scale acoustic mode causes strong dissipation \cite{suppl,FC,GFF,broadbent.79,kopev.83,melander..88,zhang....13,klyatskin66,mitchell..95,naugolnykh14}. During the time intervals when vortices stop oscillating and are comparable in magnitude, the energy continues approximately linear growth.

The evolution is different in run E, where condensate vortices do not appear, and the Mach number $M\simeq 0.6$ is reached at the scales below the box size, leading to many mid-size vortices being present at the saturated stage, which fluctuates much less as a result. Note that the energy in the left panel is normalized by pumping. As is clear from the right panel of Fig.~\ref{e-of-t}, in all runs the typical velocities and total energy reach approximately the same values. Incidentally, it was reported that even a stable condensate does not appear in optical turbulence when the pumping is too strong \cite{Cond}; whether there is a general phenomenon of inverse cascade disruption due to high effective nonlinearity (for instance, because integrals of motion cease to be quadratic) is poorly understood.

It is important that the decay and growth of total energy is determined by the shock (rather than total) dissipation; apparently, solenoidal dissipation, however large and finite the kinematic viscosity $\nu$, is irrelevant to the inverse cascade.

\begin{figure}
  \centering
  \includegraphics[scale=0.6]{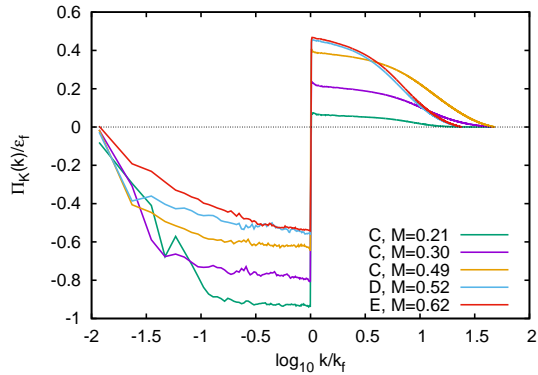}
  \caption{Kinetic energy fluxes for the case C at $M=0.21$ and 0.30 (nonstationary, no condensate) and $M=0.49$ (quasi-stationary with condensate); D at $M=0.52$ (quasi-stationary with condensate); E at $M=0.62$ (quasi-stationary, no condensate). Fluxes saturate at $\Pi_K\approx\pm0.5\varepsilon_f$ as the Mach number approaches 0.6 apparently due to a  direct acoustic energy flux that matches the energy injection rate at $M\geq0.6$, halts the total energy growth, and forms a closed energy flux loop at $k>k_f$. }
  \label{flux-all}
  \vspace{-0.2cm}
\end{figure}
\begin{figure*}
  \centering
  \includegraphics[scale=0.6]{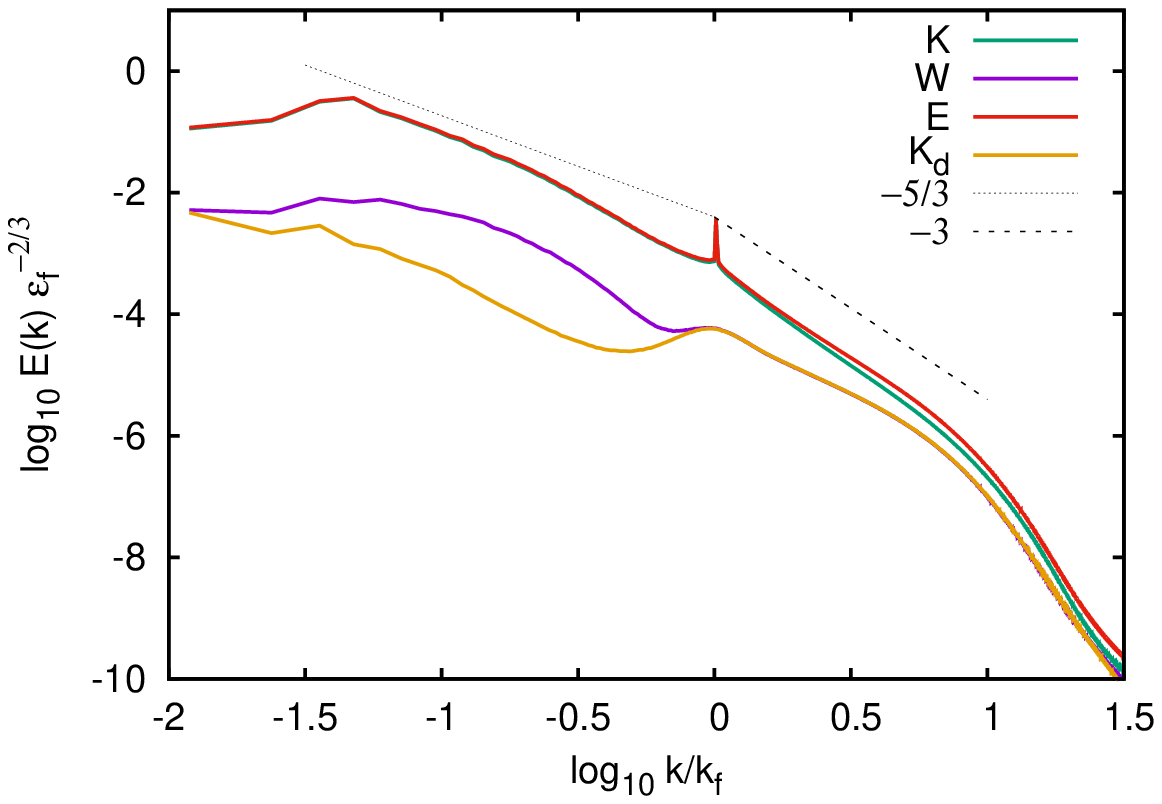}
  \includegraphics[scale=0.6]{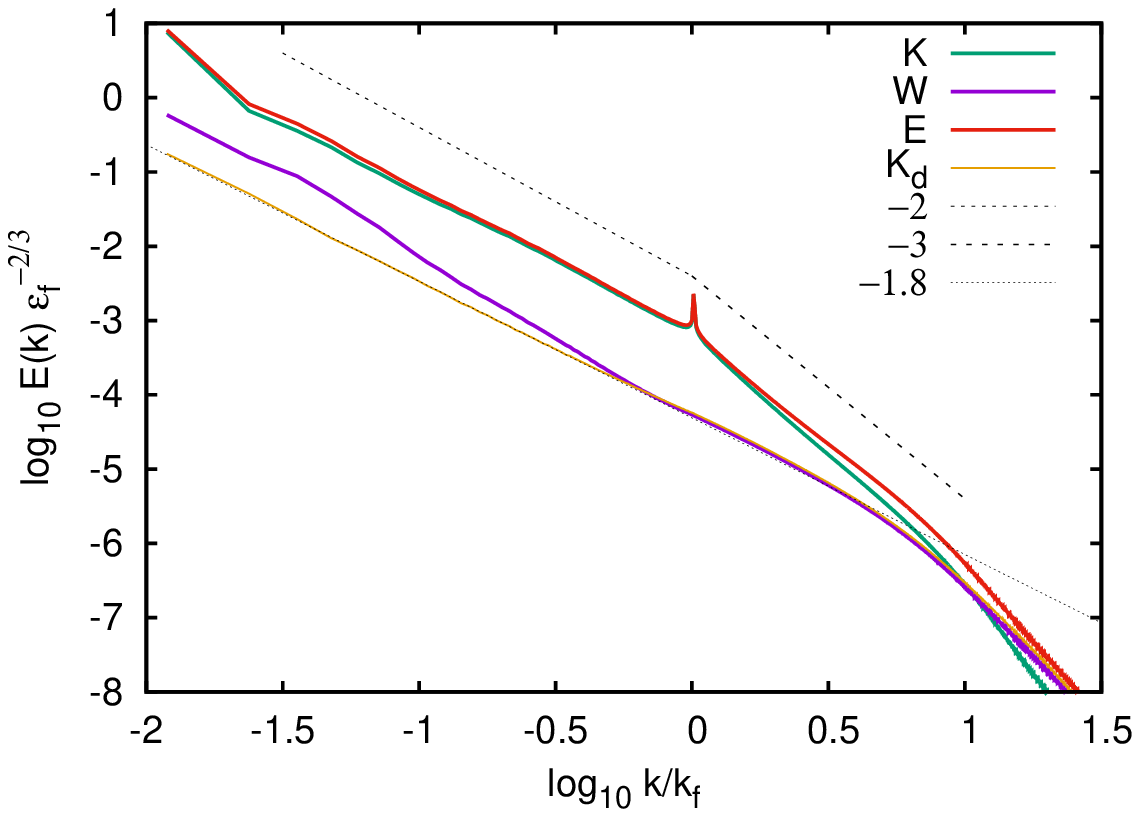}
  \caption{Energy spectra for case C at $M=0.2$ (left) and 0.5 (right); see Refs.~\cite{suppl,kida.90,sagaut,galtier.11,wang.....12} for definitions in the compressible case.}
  \label{copow}
  \vspace{-0.2cm}
\end{figure*}

Let us now analyze kinetic energy fluxes for states with different Mach numbers that appear at different pumping rates and in different boxes (Fig.~\ref{flux-all} and Ref.~\cite{suppl}).
 Consider that we use solenoidal pumping so that at a low Mach number we have practically incompressible turbulence with most of the energy going to the left of the forcing scale, i.e., into the inverse cascade. 
We see that with increasing Mach number, the larger and larger fraction of kinetic energy goes to the right of the forcing scale, i.e., into the direct cascade. Still, the inverse cascade is well pronounced in all cases. At Mach numbers of order unity, approximately equal fluxes go to large and small scales (a turbulence version of energy equipartition).
Figure~\ref{copow} shows the spectra of kinetic and potential energy and also separately the spectrum of the kinetic energy of potential (dilatational) part of the velocity field. We see that $W(k)\ll K(k)$ for most $k$, yet their fluxes must be  comparable to provide for a steady state. This is an extra reminder how different (and complementary) information is brought by analyzing together the spectra and fluxes. We also see that the dilatational part of the kinetic energy is small all the way to $k\simeq 10k_f$, so that the kinetic energy of vortices dominates. Only at $k>10k_f$ do the waves dominate and kinetic and potential energies and fluxes are getting equal, as it must be in weak turbulence \cite{ZLF}.

Turbulence at scales smaller than the force scale can be naturally assumed to carry the direct energy flux provided by acoustic waves. This is supported by the spectra, which show that at $k\gtrsim k_f$ the potential energy is equal to the dilatational energy and both decay as $k^{-2}$, as expected for acoustic turbulence \cite{ZLF}.
Of course, there is more to turbulence at $k>k_f$ than the energy cascade. It must also carry the cascade of $H$ which it does by the solenoidal part of the velocity field whose vorticity spectrum behaves as $k^{-1}\ln^{-1/3}(k/k_f)$, in exact correspondence to the theory of the enstrophy cascade in incompressible turbulence \cite{Kra,Kra71,FL}. An effective Mach number decreases towards small scales so that vortices and waves are getting decoupled. Since the vortical contribution decays faster, waves dominate kinetic energy for the small-scale part of the spectra. For most wave numbers, however, the energy of the vortices is dominant.

Let us now look at turbulence at scales above the force scale. It is likely that the acoustic direct energy cascade originates at the scales far exceeding the force scale and goes through it. This is evidenced by the spectra of potential and dilatational kinetic energy which behave continuously through $k_f$. On the contrary, the solenoidal part of the kinetic energy has a narrow peak right at $k=k_f$ and the kinetic energy flux $\Pi_K$ jumps from $\Pi_W$ at $k>k_f$ to $-\Pi_W$ at $k<k_f$, so that the total energy flux towards large scales is zero.

It deserves attention that the low-Mach energy spectra at $k<k_f$  in the left panel are usual $k^{-5/3}$,
while the spectra are close to $k^{-2}$ in the right panel of Fig.~\ref{copow}. However tempting it is to ascribe this to shock waves, this is  not the case since the spectra are overwhelmingly dominated by the kinetic energy of solenoidal motions, i.e., vortices. That means that even though density variations are substantial only at large scales (where the Mach number is not small), they modify vortices  and affect spectra at all scales down to the pumping scale. Our compressible spectrum climbs towards small $k$ faster than the $-5/3$ spectrum of an inverse cascade with a large-scale sink (yet slower than the $k^{-3}$ spectrum of the large-scale coherent vortex \cite{Shats}). To interpret this, recall that the mechanism of inverse cascade is the deformation of small vortices inside a large one and the back reaction which reinforces the large vortex \cite{Cheyink}. In a compressible case, the fact that the spectrum is steeper may mean that for a cascade to proceed, the ratio between energies of the larger vortex and smaller vortices inside it must be larger than in an incompressible case.

That potential energy exceeds dilatational energy at large scales is an extra evidence that density perturbations are related not only to waves but also to vortices. Note that a similar detailed energy equipartition across scales is seen in three dimensions (3D) at $M\approx0.6$ \cite{sarkar...91,jagannathan.16}.
\begin{figure*}
  \centering
  \includegraphics[scale=.3]{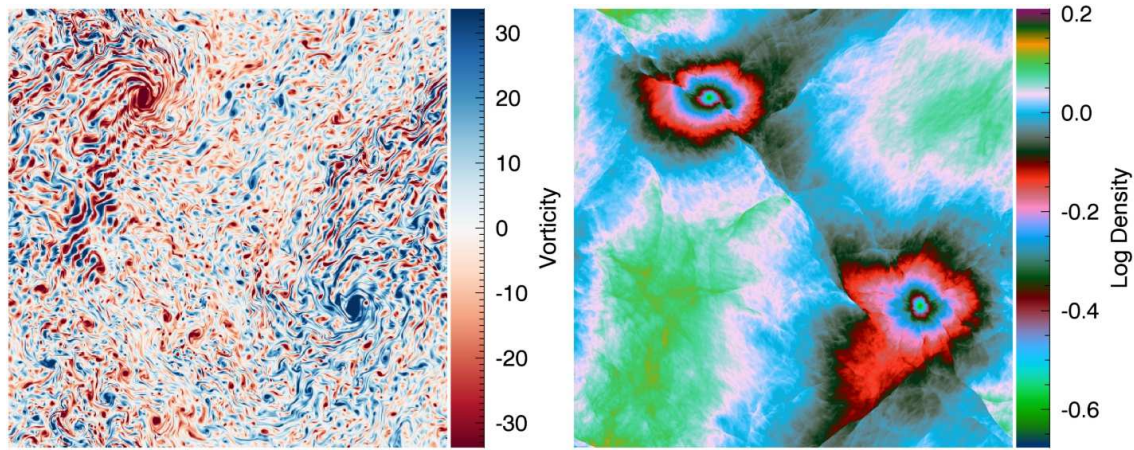}
  \caption{Vorticity and density fields in model B at $t=412$. (See Ref.~\cite{suppl} for animations and visuals generated using the line integral convolution technique \cite{cabral.93}.)
  }
  \label{plate}
  \vspace{-0.2cm}
\end{figure*}

When the inverse cascade reaches the system size and creates coherent vortices, their main dissipation is at the shocks, which go out of the vortex centers, connect them, and create spiral structures around them.
Apart from a purely fluid-mechanical interest, the spontaneous formation of strong vortices with shocks in compressible quasi-2D flows may influence different astrophysical phenomena, for instance, in the contexts of disk accretion \cite{gammie},  galactic disks \cite{bournaud}, or planetesimal formation in protoplanetary disks \cite{marcus......15}. 
Here, we focus on analyzing the vortex structure and the energy-momentum fluxes that support the coherent vortex. To appreciate better the peculiarities of the compressible case, let us briefly remind that in the incompressible case the inverse cascade produces a pair of vortices with a narrow viscous core, outside of which the mean azimuthal velocity is independent of the radius, $U=\sqrt{3\varepsilon_f/\alpha\rho }$, where $\alpha$ is the rate of uniform (bottom) friction \cite{laurie14}. Each vortex is sustained by the inward radial momentum fluxes. The (radial) flux of the radial momentum is provided mostly by the mean pressure, $\rho U^2\approx rdP/dr$.  The flux of the azimuthal momentum is provided by fluctuations, $\tau=\rho \langle uv\rangle=r\sqrt{\rho\varepsilon_f\alpha/3}$, where $u,v$ are respectively azimuthal and radial fluctuating velocities in a reference frame comoving with the vortex center. For the flat profile, the turbulence-vortex energy exchange rate per unit area, $F_1=r^{-1}\partial_rrU\tau=2\varepsilon_f$,  is also independent of the radius and equal to twice the input rate; the turbulence flux divergence $F_2=r^{-1}\partial_rr\langle v(\rho u^2+\rho v^2+2p)\rangle/2$ is negligible \cite{laurie14}. All the energy input from the external pumping and turbulence inverse cascade is dissipated inside the vortex by linear friction, $\alpha \rho U^2=3\varepsilon_f$.

\begin{figure*}
  \centering
  \includegraphics[scale=0.475]{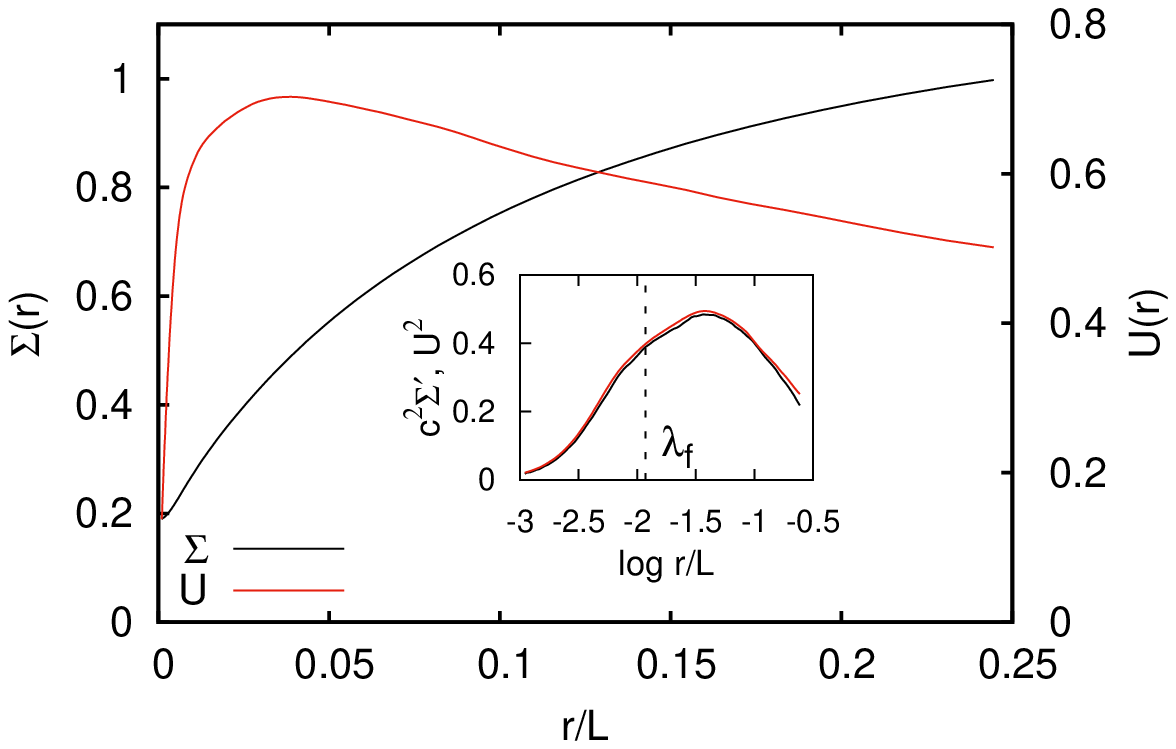}
  \includegraphics[scale=0.45]{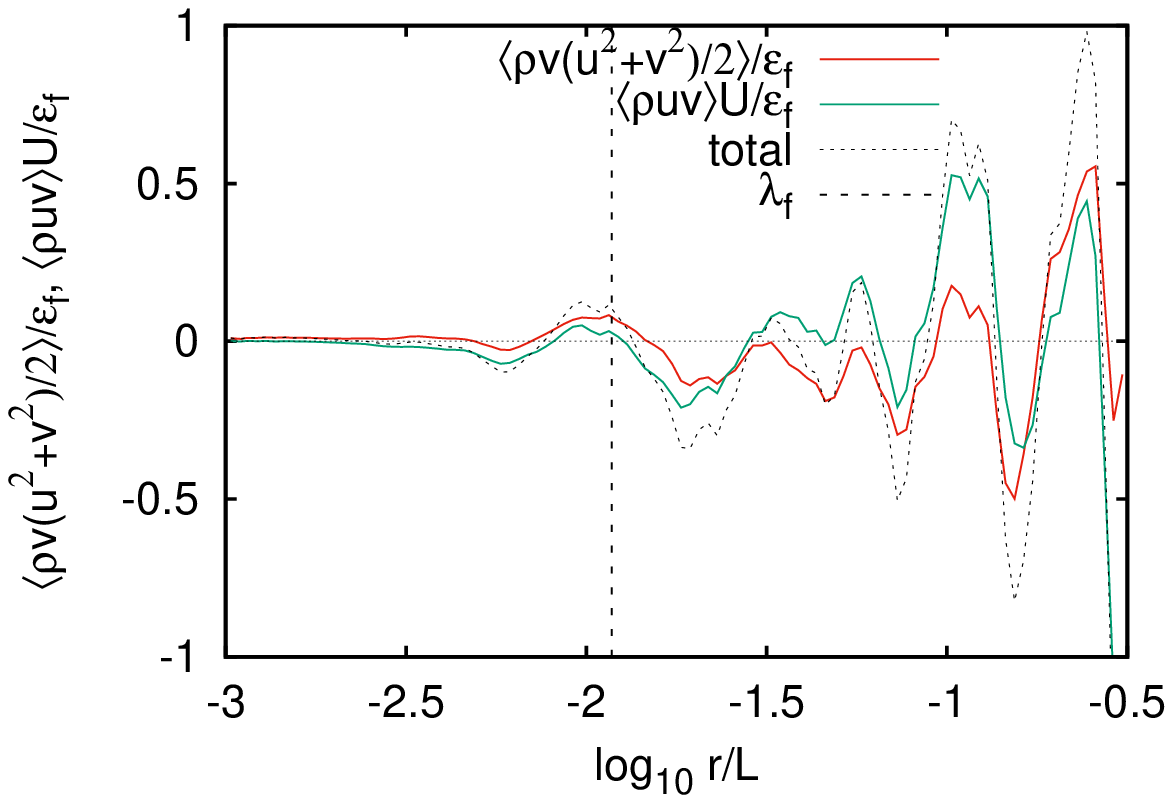}
  \includegraphics[scale=0.45]{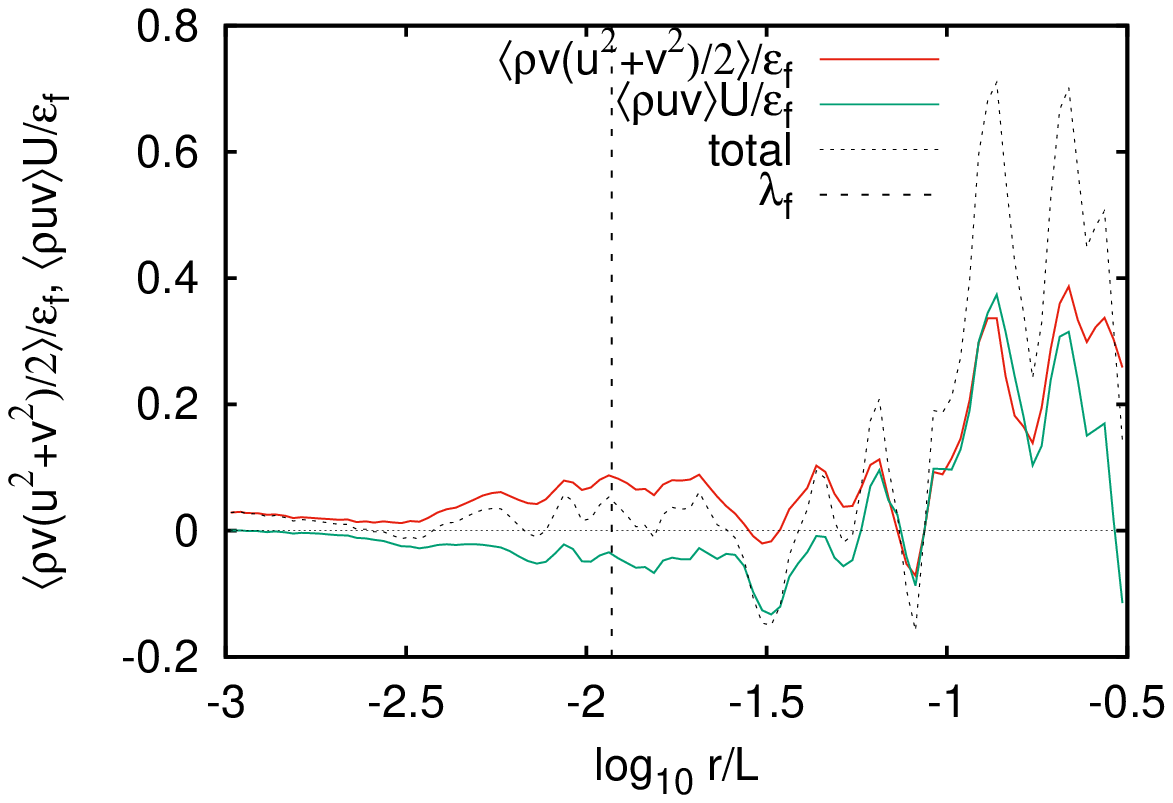}
  \caption{Profiles of the mean density $\Sigma(r)$ and velocity $U(r)$  for condensate in model C at $M\approx0.5$, $t\in[380,400]$ (left panel). The inset illustrates the radial force balance for the mean flow. Time-averaged turbulent fluxes for case C at $t\in[283,300]$ (center) and $t\in[380,400]$ (right panel). The sum of two fluxes is shown by the dashed line; vertical lines indicate the forcing scale $\lambda_f$.
  }
  \label{dpro}
  \vspace{-0.2cm}
\end{figure*}

For the compressible case which we consider, there is no bottom friction and, consequently, no momentum loss from the system. Averaging the continuity equation for the density of angular momentum and taking into account that the mass flux $\langle\rho v\rangle$  must be zero, we get inside the vortex the condition for total zero momentum flux,
\begin{eqnarray}
\langle\rho vu\rangle= 
\nu\Sigma r \frac{\partial}{\partial r}{U\over r}+\left\langle{\nu\rho\over r}{\partial v\over \partial\phi}\right\rangle+ r\left\langle\nu\rho{\partial\over\partial r}{u\over r} \right\rangle.\label{comp2}
\end{eqnarray}
Since the mean density and velocity profiles $\Sigma(r)$ and $U(r)$ are smooth, then the first term on the right-hand side (rhs) goes to zero with $\nu$. It is thus clear that inertial momentum flux $\tau=\langle\rho uv\rangle$ could be nonzero only if there is a finite inviscid limit of the last two terms on the rhs, which are due to turbulence. That requires tangential discontinuities, apparently provided  by the spiral shocks coming out of the vortices and long shocks connecting vortex centers which we observe. Movies \cite{suppl} also show density profiles strongly corrugated along $\phi$ and fast changing along $r$, which is conducive to having $\tau\ne0$. Most likely, the mechanism of nonzero viscous momentum transfer at the inviscid limit is the angle change of a streamline after passing through shock. For that one indeed needs spiral-like shocks, which deflect streamlines out. If the shock deflects against  the vortex flow, then the rhs of (\ref{comp2}) is   negative, as can be expected (inertia brings momentum in, and viscous friction takes it out).   Only with nonzero inertial momentum flux into a vortex can the inertial energy flux into vortex $\tau r\partial_r(U/r)$ be nonzero as well. We expect the energy input rate $\varepsilon_f$ to exceed the viscous energy dissipation outside the vortices, so that the energy must flow into the vortices to be dissipated there.

A typical snapshot of the vorticity and density fields in the energy-saturated state with a condensate is shown in Fig.~\ref{plate}. To compute the averages, we save 20 flow snapshots per crossing time; with these we can create reasonably smooth animations \cite{suppl} and robustly decompose the mean flow from the turbulence, using an algorithm similar to that of Ref.~\cite{laurie14} with some modifications to account for compressibility \cite{suppl}. One learns from case C that the condensate vortex has a circular core with a vorticity comparable to several other vortices present at any given time; what distinguishes the condensate vortex is a spiral around the core so that the density is perturbed in the whole region, core and spiral, which provides for strong dissipation. Secondary vortices, on the contrary,  do not perturb density in any substantial way. 
Most of the time, the vortex as a whole moves generally with a speed much less than the flow velocity in the vortex itself, so one can neglect distortions caused by the center motion.

As in Ref.~\cite{laurie14}, the mean vorticity profile within a coherent vortex is close to isotropic, but the 2D density distribution shows shallow diagonal minima, resembling the density depressions along the lines connecting the vortices in individual snapshots (Fig.~\ref{plate}). The mean flow is characterized by the azimuthally averaged density and velocity profiles $\Sigma(r)$ and $U(r)$, shown in the left panel of Fig.~\ref{dpro} for case C. We see that  the density decreases monotonically towards the center, while the velocity grows and then decays. As the vortex grows, the outer velocity profile flattens.  What matters, however, is the comparison of the centrifugal force $\Sigma U^2/r$ and the radial pressure gradient $dP/dr=c^2d\Sigma/dr$.  It is more convenient to compare $U^2$ with $c^2d\ln\Sigma/d\ln r\equiv c^2\Sigma'$, which is done in the inset in Fig.~\ref{dpro}. Remarkably, the radial balance of the momentum fluxes holds with an accuracy of a few percent already on the mean profiles, despite the quite complicated structure of the vortex, as seen in Figs.~\ref{plate} and \ref{dpro}. In other words, the mean pressure and velocity satisfy the steady Euler equation, that is, the contribution of fluctuations into the radial momentum flux is negligible, even though the fluctuations are quite strong (and $U_{\rm max}\approx0.7c$).

Fluctuations, however, play a crucial role in feeding energy and azimuthal momentum to the vortex, as shown in Fig.~\ref{dpro}. 
We find that the fluxes here are quite different from the incompressible case: $F_1, F_2$ are comparable but they change sign along the radius (see Fig.~\ref{dpro} and Ref.~\cite{suppl}). Apparently, those oscillations are a signature of the spiral structure of shocks, so that the energy fluxes converge to shocks at spiral arms rather than to the vortex center.

Note that fluxes fluctuate strongly, so that a short-time average can often give an opposite sign of the energy fluxes, as seen from comparing the two right panels in Fig.~\ref{dpro}. A positive (outward) angular momentum flux, as that observed here at some radii at the vortex periphery, was previously observed in the case of a rotating disk and ascribed to compressibility \cite{gammie}.

\begin{acknowledgments}
This research was supported in part by the National Science Foundation through Grant No.~AST-1412271 as well as through
XSEDE allocations on {\em Trestles}, {\em Gordon}, {\em Comet} supercomputers at SDSC and {\em Stampede} at TACC (projects MCA07S014 and SDSC-DDP189). The work of G.F.  
was supported by the RScF Grant No.~14-22-00259 and by the ISF grant No.~712174. The authors gratefully acknowledge support from the Simons Center for Geometry and Physics, Stony Brook University, at which some of the research for this paper was performed. A.K. appreciates the warm hospitality of Weizmann Institute of Science during his visit in late 2016.
\end{acknowledgments}

 {

\section{Supplemental Material}
\subsection{A. Description of numerics}
\vspace{-.2cm}
\setlength{\tabcolsep}{0.16cm}
\begin{table*}
	\caption{Simulations and parameters.
	    {\small
$N=N_x=N_y$ -- linear grid resolution;
$\lambda_f=2\pi/k_f$ -- energy injection length scale;
$\varepsilon_f$ -- energy injection rate per unit area;
$\varepsilon_g$ -- actual maximum total energy growth rate with established inverse cascade;
$\tau_f=\rho_0^{1/3}\lambda_f^{2/3}\varepsilon_f^{-1/3}$ -- characteristic forcing time;
$t_{\rm cnd}$ -- approximate condensation time; $t_{\rm sat}$ -- approximate energy saturation time; 
$t_{\rm end}$ -- simulation stop time; $M_{\rm end}$ -- rms Mach number at $t_{\rm end}$. 
}}\vspace{0.2cm}
	\begin{tabular}{c c c c c c c c c c c}
		\hline
		\hline
		Key & $N$ &$\lambda_f$  & $\varepsilon_f$ & $\varepsilon_g/\varepsilon_f$ & $\tau_f$& $t_{\rm cnd}$& $t_{\rm sat}$& $t_{\rm end}$ & $M_{\rm end}$\\
		\hline
		A & $512$  & 0.047 & $0.001 $ & 0.85 & 1.30&80&250&1500 & 0.54 \\
		B &$2048$ &0.023 & $0.001 $ & 0.83 & 0.82&100&280& 500 & 0.54 \\
		C &$8192$ &0.012 & $0.001 $ & 0.92 & 0.52&150&390& 450 & 0.52\\
		D &$4096$ &0.012 & $0.002 $ & 0.96 & 0.41&200&150& 250 & 0.52\\
		E &$4096$ &0.012 & $0.008 $ & 0.95 & 0.26&---&80& 200 & 0.62\\
		\hline
		\label{para}
	\end{tabular}
\end{table*}

We carried out a set of implicit large eddy simulations (ILES \cite{sytine....00,grinstein..07}) in a square periodic domain of size $L\times L$,
covered with a uniform Cartesian grid of $N\times N$ cells. The compressible Euler equations 
 {
\begin{eqnarray}
\partial_t \rho+\bm\nabla\cdot(\rho\bm u)& =&0,\\
\partial_t (\rho \bm u)+{\bf \nabla\cdot}\left(\rho\bm u\bm u + p{\bf I}\right)&=&\bm f,\label{mome}\\
\partial_t {\cal E}+{\bf \nabla}\cdot\left[\left({\cal E}+p\right)\bm u\right]& = &{\bm u\cdot \bm f},\label{ener}
\end{eqnarray}
where $\rho$ is the density, $\bm u$ -- velocity, $p$ -- pressure, and ${\cal E}=\rho (u^2/2+e)$ -- total energy density,}
were numerically solved using an implementation of the piecewise parabolic method (PPM) \cite{ppm} in the Enzo code \cite{enzo}. 
A purely solenoidal, white-in-time random external force per unit mass $\bm a=\bm f/\rho$ has been applied at an intermediate pumping scale $\lambda_f=2\pi k_f$ (such that $L/N\ll\lambda_f\ll L$). The system of conservation laws includes the energy equation (\ref{ener}) and an ideal gas equation of state  {$p=(\gamma-1)\rho e$} is assumed with an adiabatic index $\gamma=1.001$ to mimic an isothermal fluid with compression factors in shocks proportional to $M^2$.
The dimensionless units are chosen so that  the box size $L=1$, the mean density $\rho_0=1$, and the speed of sound $c\approx1$. The force supplies kinetic energy to the system at a relatively small rate $\varepsilon_f\in[0.001,0.01]$.  {Substantially higher forcing rates would inhibit the inverse cascade, as most of the injected energy would be dissipated in shocks right at the injection scale.} The force correlation time is $\sim10^{4}$ times shorter than the characteristic vortex turn-over time at the force scale, $\tau_f=\rho_0^{1/3}\lambda_f^{2/3}\varepsilon_f^{-1/3}$. Each numerical model is thus fully
defined by the three input parameters $(N,\varepsilon_f, \lambda_f)$. A summary of parameters for the five computed cases is provided in Table~I. 

In the initial three-step A-B-C sequence of models, $N$ changes by a factor of four and the extent of both forward and inverse cascades increase by a factor of two per step. 
Complementary cases D and E explore the dependence on the pumping rate $\varepsilon_f$ at moderately high resolution.

Depending on the spatial resolution, the models were evolved for $t_{\rm end}=(0.2-1.5)\times10^3$ sound crossing times, starting from a uniform-density fluid at rest. Since some of the energy is dissipated at small scales, the actual maximum energy growth rate with established inverse cascade, $\varepsilon_g$, varies within $(0.83-0.96)\varepsilon_f$, depending on $\lambda_f$ and $N$.
Typical peak Mach numbers reached as the systems saturate are  subsonic yet of order unity: $M\in[0.5,0.7]$. The condensation time $t_{\rm cnd}$ is (somewhat arbitrarily) defined as the time of the last merger of two like-sign large vortices, seeding the system-size vortex dipole, which then gets further inflated with time as it continues to receive more energy. An increase in $\lambda_f$  shortens the condensation time (cases C-B-A); a sufficiently large $\varepsilon_f$ may prevent condensation (case E); insufficiently resolved forcing scale ($k_{\rm max}/k_f\lesssim20$) reduces the energy growth rate $\varepsilon_g$ and suppresses the formation of coherent vortices at scales above $\lambda_f$.

\subsection{B. Spectral energy densities and fluxes}
\vspace{-.3cm}
The total energy conserved by smooth solutions of the ideal isothermal Euler system includes kinetic and internal (potential) parts, $E=K+W$, respectively.
 {At sufficiently small Mach numbers, the kinetic energy can be further decomposed into solenoidal and dilatational parts $K\approx\rho_0({\cal K}_s+{\cal K}_d)$, using the Helmholtz decomposition of the velocity field, $\bm u=\bm u_s+\bm u_d$ ($\bm\nabla\cdot\bm u_s=0$ and $\bm\nabla\times\bm u_d=0$ \cite{sagaut}).  }

The spectral densities shown in Fig.~3 are defined by $K(k)=P(\rho\bm u,\bm u;k)/2$ and $W(k)=P(\rho,e;k)/2$. % \cite{BK16}.
Here the kinetic energy spectral density is represented by a cospectrum $P(\bm j,\bm u;k)$ of the momentum density $\bm j\equiv\rho\bm u$ and velocity $\bm u$, i.e. the Fourier transform of the symmetric part of the cross-covariance function, $\Gamma^s_{ju}(\bm r)=\langle\bm j\bm\cdot\bm u'+\bm j'\bm\cdot\bm u\rangle/2$, integrated over annuli on the $k$-plane, $P(\bm j,\bm u;k)\equiv\int\widehat{\Gamma^s_{ju}}(\bm \kappa)\delta(k-|\bm \kappa|)d\bm \kappa$. 
Likewise, the cospectrum $P(\rho,e;k)$ is the Fourier transform of $\Gamma^s_{\rho e}(\bm r)=\langle(\rho-\rho_0)(e'-e_0)+(\rho'-\rho_0)(e-e_0)\rangle/2$, integrated over annuli in $k$-space. We use angular brackets to denote ensemble averages over point pairs separated by the lag $\bm r$, while $\rho_0$ and $e_0$ denote average density and average specific potential energy in the domain. 

As soon as the spectral densities $K(k)$ and $W(k)$ are defined, one can assume homogeneity and use a point-split version of the energy balance equation for a forced isothermal Euler system to determine the kinetic energy transfer function in a compact symmetric cross-covariant form ${\cal T}_K(\bm r) = -\frac{1}{2}(\Gamma^{\rm s}_{\rm j, (\rm u\cdot\nabla)\rm u}+ \Gamma^{\rm s}_{\rm u, (\nabla\cdot\rm j)\uu+(\rm j \cdot \nabla)\rm u})$. The spectral flux of kinetic energy is then defined in a standard way by $\Pi_K(k)=\int_k^{\infty}T_K(\kappa)d\kappa$, where the kinetic energy spectral transfer function $T_K(k)=\int\widehat{{\cal T}_K}(\bm\kappa)\delta(k-|\bm \kappa|)d\bm\kappa$.%, see \cite{BK16} for a detailed derivation.
 The definitions  of kinetic energy spectra and fluxes adopted here generally follow \cite{galtier.11} and differ from \cite{kida.90}, where point splitting is done in a symmetric way, generalizing the incompressible formulation by replacing $\bm u$ with $\bm w\equiv\sqrt{\rho}\bm u$. In compressible turbulence at $M\lesssim1$, the specific form of density weighting is not very important, as density variations remain small and velocity spectra almost overlap with spectra of $\bm w$ \cite{wang.....12}. 
 The dilatational part of the kinetic energy shown in Fig.~3 is reasonably well approximated by $K_d(k)\approx \rho_0{\cal K}_d(k)=\rho_0k^{-2}P(\bm\nabla\cdot\bm u,k)/2$ \cite{sagaut}. Our definition of $W(k)$ is based on a cross-covariance function that differs from one adopted in \cite{galtier.11} by a factor of $1/2$. This is needed to ensure equipartition of dilatational kinetic energy and potential energy $K_d(k)\approx W(k)$ in the acoustic limit, which is observed in Fig.~3 at $k>k_f$; see discussion of the subject in \cite{sarkar...91,sagaut,jagannathan.16}.

\begin{figure*}
	\centering
	\includegraphics[scale=.885]{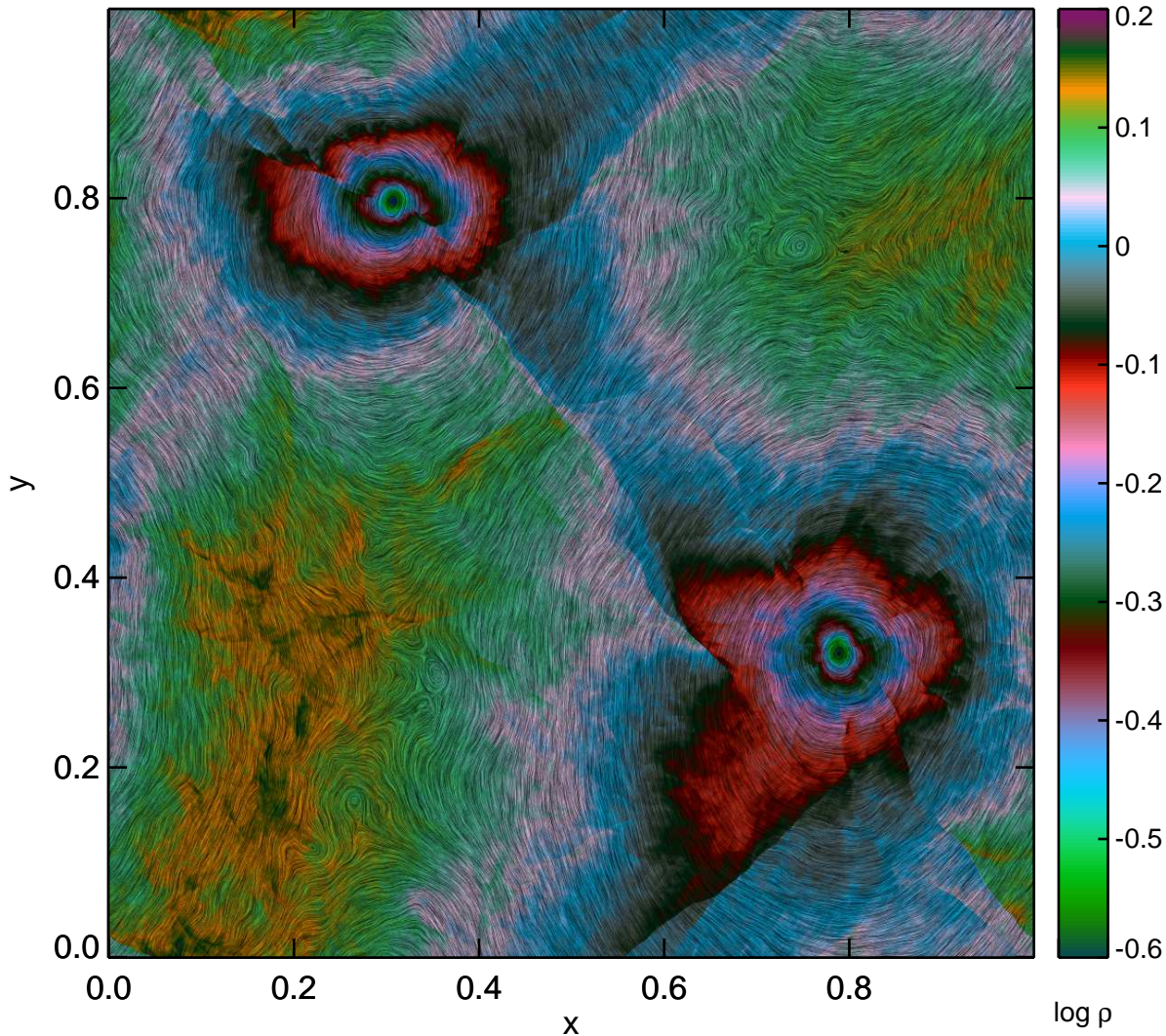}
	\caption{Large-scale flow structure with condensate and associated large-scale shock on top of smaller scale turbulence for a case B snapshot at $t=412$ (same as in Fig.~4 of the main text). The drapery texture highlights fluid streamlines, while color indicates the density structure, including shocks and shocklets. The upper-left and lower-right vortices, constituting the large-scale dipole, rotate clock-wise and counter-clock-wise, respectively.}
	\label{contur}
\end{figure*}

\subsection{C. Acoustic destabilization of vortices and intermittency}
%\vspace{-.3cm}
%
It deserves mentioning that temporally the flux loop is very intermittent: long periods of energy growth at large scales are interspersed with relatively short bursts of the energy transfer to small scales, thus expressing the irreversibility of turbulence statistics \cite{FC,GFF}. Individual vortices are destabilized due to emission of sound waves \citep{broadbent.79,kopev.83}. When a sufficiently close pair of equal vortices leapfrog and merge \citep{melander..88}, strong acoustic noise is generated. Unequal co-rotating vortices emit sound from leapfrogging, shearing and tearing \citep{zhang....13}. Loose vortex pairs with parallel spins will slowly diverge \citep{klyatskin66}, as they radiate sound \citep{mitchell..95,zhang....13}. Vortex mergers trigger significant pressure variations in the near field and acoustic emission in the far field. As acoustic losses tighten a dipole, it accelerates to sonic velocities, producing strong shocks. Strong acoustic emission can trigger dipole collapse \citep{naugolnykh14}.

\begin{figure*}
	\centering
	\includegraphics[scale=.9]{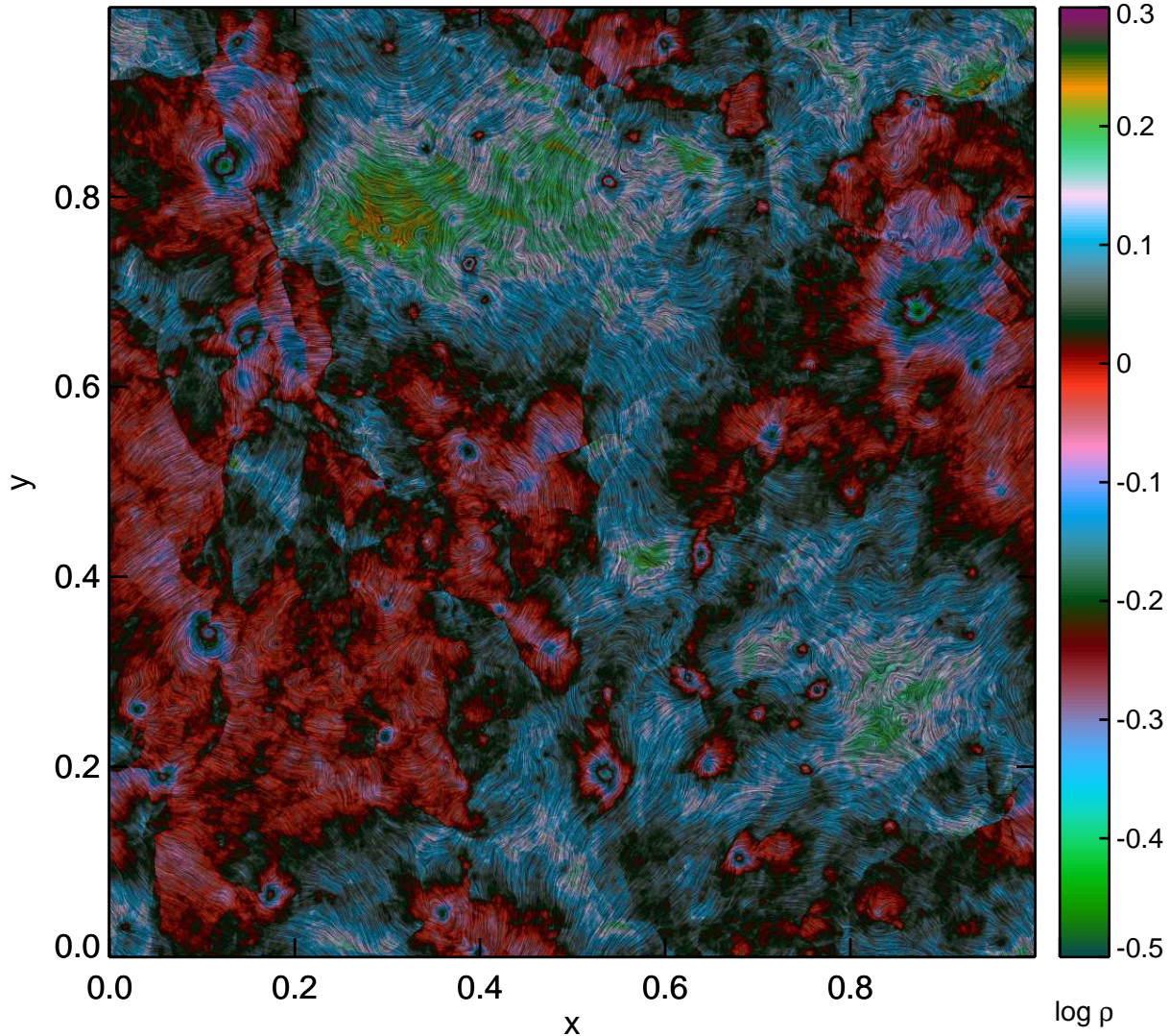}
	\caption{Same as Fig.~\ref{contur}, but for a case E snapshot at $t=200$. At high pumping rate, the flow fields get more intricate and the condensate takes a form of two counter-rotating large-scale clusters of (primarily) like-sign vortices. A careful inspection of the streamlines' texture, obtained after filtering out small-scale structures, allows identification of two distinct clusters located in the upper-right and lower-left parts of the domain.}
	\label{clusters}
\end{figure*}

\subsection{D. The structure of energy condensate}
A variety of techniques can be used to highlight distinct features of turbulent flows through richly informative and spectacular images. 
Here we apply the line integral convolution (LIC) algorithm developed in \cite{cabral.93} to visualize the velocity vector field associated with energy condensate, using drapery texture generation, and then color the resulting texture based on the local density values.

This technique combines velocity and density information together in a single visual (e.g., Fig.~\ref{contur}), which is complementary to a more traditional way of presenting results in the form of separate vorticity and density plots (e.g., Fig.~4 of the main text). This seems to be the first application of the LIC to simulations of 2D turbulence.

Note that LIC involves filtering out small-scale turbulent structures which helps to identify coherent vortices at intermediate scales between the forcing and energy condensate scales. Such eddies, clearly visible in Fig.~\ref{contur}, can be responsible for intermittency in the inverse energy cascade. Fig.~\ref{clusters} further illustrates clustering of like-sign vortices forming the energy condensate in case E with high pumping rate.

\subsection{E. Vortex tracking and profiling}
To decompose the mean flow controlled by the box-scale vortex dipole from turbulence, we use an algorithm similar to that of \cite{laurie14}, with some modifications to account for compressibility. As in \cite{laurie14}, our approach is based on stacking of a large number of vortex realizations (using 20 flow snapshots per box-crossing time and averaging over $15-50$ crossing times) to derive mean velocity and density profiles.
For each flow snapshot, we locate the center of the positive vortex as the maximum of potential vorticity and then compute the center-of-mass of $\omega/\rho$ in a box of $9\times9$ cells around the maximum. We then center the box around the vortex and transform the velocities to a reference frame co-moving with the center. We obtain the mean rotation velocity and density profiles $U(r)$ and $\Sigma(r)$ by stacking a series of snapshots together and thus filtering out the zero-mean fluctuations. Repeating the same procedure for the negative vortex and accounting for the opposite vorticity sign, we further double the number of realizations. The vortex detection algorithm described above is robust at low resolution and at low Mach numbers, when shocks are weak and only two strong vortices are present. In case C, however, we had to smooth the potential vorticity field with a $4\times4$ boxcar average to capture vortex centers correctly. Without this critical step, the algorithm would intermittently capture strong oblique shock fronts instead of vortex centers, resulting in $\sim4$\% misidentifications that spoil the mean profiles of turbulent fluxes in the core region. As the condensate evolves toward a stationary state in our high resolution case C, a number of medium-size coherent vortices present along with the box-size dipole. These objects were also able to confuse the algorithm, when a tight secondary vortex core made a close passage to the primary of the same sign. Animations were instrumental in helping us to control the algorithm and discard misidentifications ($\sim0.4$\% of cases where the secondary was erroneously recognized as primary) from the stack of snapshots used to derive the mean profiles. Overall, our algorithm is stable and thus the obtained mean profiles are robust. However, to make sure the mean vortex is circularly symmetric, averaging over at least 10 turnover times is required.

Note that for ideal  isothermal gas with the potential energy density $W(\rho)=c^2\rho\ln(\rho/\rho_0)$, the dynamical viscosity $\eta=\nu\rho$ must be constant everywhere, since $\nu\propto 1/\rho$ at constant temperature. Would that be the case, the viscous terms could not contribute to the mean momentum balance [see equation (1) in the main text] and the turbulent momentum flux into the vortex must also be zero in a steady state. In our case, the kinematic viscosity is determined by the grid discreteness, so that the effective dynamic viscosity is not exactly constant and the turbulent momentum flux is non-zero, but small.

\subsection{F. Isothermal approximation for moving soap films}
In the leading-order approximation, the dynamics of thin soap films can be described by the compressible Euler equations, if the fluid viscosity and the surfactant solubility can be neglected and fluid velocities are of order of the Marangoni elastic wave speed \cite{Chomaz01}. In this case, soap films behave as an isothermal two-dimensional gas with $\gamma=1$. The equations governing film dynamics contain the film thickness $h$ (replacing the density $\rho$), the in-plane velocity $\bm u$ locally averaged over $h$, average surfactant concentration in the interstitial liquid, and surface surfactant concentration. The isothermal fluid approximation applies when non-diffusive soap is considered, assuming initial variations of the total soap concentration are negligible. In addition, it is required that the bending and elastic Mach numbers are of the same order, $M_b\sim M_e$ \cite{Chomaz01}.

\subsection{G. Near-isothermal conditions in applications to interstellar turbulence}
In the interstellar medium (ISM), balance of heating and cooling processes plays an important role in setting the temperature of the gas \cite{wolfire...03}. Heating is provided by cosmic rays and by the interstellar radiation field (including photoionization of atomic carbon, far ultraviolet photoelectric heating on small dust grains and large molecules, irradiation of interstellar dust, and stellar X-rays). Cooling mostly occurs through emission lines and continuum radiation, some or all of which escapes the region and carries away energy. Radiative cooling processes include: collisionally excited atomic line cooling in the diffuse ISM, fine structure line cooling by atomic oxygen and ionized carbon in cold ($\sim100$~K) clouds of atomic hydrogen, molecular cooling (e.g., CO rotational line emission), thermal and non-thermal emission by dust grains in cold dense molecular clouds ($\sim15$~K). The details of thermal balance depend on radiative transfer and couple with very complex interstellar chemistry. The expected thermal equilibrium is linearly unstable with respect to isobaric perturbations at temperatures between $\sim100$~K and $\sim6000$~K \cite{field65}. This defines `warm' and `cold' stable phases with $T\sim6000-12,000$~K and $T\sim80-100$~K, densities $\sim1$~cm$^{-3}$ and  $\sim100$~cm$^{-3}$, respectively. Dense ($10^{3-6}$~cm$^{-3}$) molecular clouds represent another phase with $T\sim10-20$~K. In some more exotic cases, isentropic modes get destabilized as well, resulting in overstability of acoustic waves \cite{field65,kritsuk86,krasnobaev.17}. Overall, the ISM is very dynamic and strong compressible turbulence and magnetic fields effectively keep it out of equilibrium \cite{kritsuk..17}. Historically, however, simple (near) isothermal models were widely used  and proved quite helpful in early studies of star formation (see \cite{mckee.07} for a review) and global ISM dynamics in galactic disks (e.g. \cite{martos.98}).

\end{document}